\documentclass[letterpaper,prl,reprint,preprintnumbers,notitlepage]{revtex4-1}

\usepackage{graphicx,color}
\usepackage{amssymb,amsmath}

\makeatletter

\begin{document}

\preprint{UTHEP-712}

\title{Scattering amplitude from Bethe-Salpeter wave function inside the interaction range}

\date{
\today
}

\author{Yusuke~Namekawa}
\affiliation{Faculty of Pure and Applied Sciences,
             University of Tsukuba, Tsukuba, Ibaraki 305-8571, Japan}
\author{Takeshi~Yamazaki}
\affiliation{Faculty of Pure and Applied Sciences,
             University of Tsukuba, Tsukuba, Ibaraki 305-8571, Japan}
\affiliation{Center for Computational Sciences, University of Tsukuba,
	     Tsukuba, Ibaraki 305-8577, Japan}
\affiliation{RIKEN Advanced Institute for Computational Science,
             Kobe, Hyogo 650-0047, Japan}


\begin{abstract}
We propose a method to calculate
scattering amplitudes
using the Bethe-Salpeter wave function
inside
the interaction range
on the lattice.
For an exploratory study of this method, 
we evaluate a scattering length of $I=2$ S-wave two pions
by the use of the on-shell scattering amplitude.
Our result is confirmed to be consistent with the value
obtained from the conventional finite volume method.
The half-off-shell scattering amplitude
is also
evaluated.
\end{abstract}

\maketitle

\setcounter{equation}{0}


{\bf \em Introduction:}
Calculation of hadronic interactions
by
lattice QCD 
is an important direction
toward understanding
fundamental properties of hadrons from 
the first principle
of the strong interaction.
In many lattice studies of
two hadrons,
the scattering phase shift $\delta(k)$
or the
scattering length $a_0$
was
evaluated
using
the finite volume method
proposed by L\"uscher~\cite{Luscher:1986pf,Luscher:1990ux}.
Energy
eigenvalues of
two hadrons on a finite volume
are related to
$\delta(k)$
in the infinite volume
through a known function.
This relation was derived from a two-particle wave function 
in (relativistic) quantum mechanics~\cite{Luscher:1990ux}
and also from the Bethe-Salpeter (BS) wave function
in quantum field theory~\cite{Lin:2001ek,Aoki:2005uf}.
In both
cases, the derivation
utilized
wave functions outside the interaction range $R$ of the two particles.
In contrast,
a relation between
the on-shell scattering amplitude and
the BS wave function inside $R$ was discussed in quantum field theory
in the infinite volume
~\cite{Lin:2001ek,Aoki:2005uf,Yamazaki:2017gjl}.
A method using a potential from the BS wave function
was also proposed
based on quantum mechanics~\cite{Aoki:2009ji}.

In this paper, extending the
quantum field theoretical
discussion in the infinite volume, 
we propose a method to calculate 
the on-shell and half-off-shell scattering amplitudes using
the BS wave function inside $R$
on a finite
volume
lattice.
We perform a simulation in quenched QCD at
the pion mass
$m_\pi = 0.86$ GeV
to evaluate the scattering amplitudes of
the isospin $I=2$ S-wave two-pion scattering in the center-of-mass frame.
Using the on-shell amplitude, we investigate
the consistency of our method
with the finite volume method by examining a
condition
of
the finite volume method
and by comparing $\delta(k)$ directly.
We also demonstrate that the half-off-shell
scattering amplitude can be calculated in a similar way.
We attempt
to extract information of the scattering from 
the half-off-shell scattering amplitude.


\vspace*{2mm}
{\bf \em Formulation:}
The BS wave function
of
two pions in the infinite volume $\phi_\infty({\bf x};k)$ 
is related to 
the scattering amplitude~\cite{Lin:2001ek,Aoki:2005uf,Yamazaki:2017gjl},
\begin{eqnarray}
 \phi_\infty({\bf x};k)
 &=& e^{i {\bf k} \cdot {\bf x}}
 + \int \frac{d^3 p}{(2 \pi)^3}
        \frac{H(p;k)}{p^2 - k^2 - i \epsilon} e^{i {\bf p} \cdot {\bf x}},
\label{def:phi_inf}
\end{eqnarray}
%
%
where $H(p;k) = \frac{E_p + E_k}{4 E_p E_k} M(p;k)$ with 
$M(p;k)$ being the half-off-shell scattering amplitude
defined by the Lehmann-Symanzik-Zimmermann(LSZ) reduction formula~\cite{Lin:2001ek,Aoki:2005uf}
and $E_k = 2 \sqrt{m_\pi^2 + k^2}$.
Some overall factors in the expression for
$\phi_\infty({\bf x};k)$ are omitted.
We consider only the S-wave scattering
in the center-of-mass frame
and neglect the inelastic scattering contribution.
At on-shell $p=k$, 
$H(k;k)$ is written by the scattering phase shift $\delta(k)$ as
\begin{eqnarray}
 H(k;k) = \frac{4 \pi}{k} e^{i \delta(k)} \sin \delta(k).
 \label{eqn:H_kk_infinite_volume}
\end{eqnarray}
In the following, $H(p;k)$ is called
the scattering amplitude
for simplicity, though its normalization differs from $M(p;k)$.
The reduced BS wave function $h_\infty({\bf x};k)$
is defined by
$\phi_\infty({\bf x};k)$ as in Refs.~\cite{Aoki:2005uf,Yamazaki:2017gjl},
\begin{eqnarray}
 h_\infty({\bf x};k)
 &=& (\sum_{i=1}^{3} \partial_i^2 + k^2) \phi_\infty({\bf x};k)
 \nonumber
 \\
 &=& - \int \frac{d^3 p}{(2 \pi)^3} \, H(p;k) e^{i {\bf p} \cdot {\bf x}} ,
\end{eqnarray}
where
we use
Eq.~(\ref{def:phi_inf}) in the last equality.
It is assumed that $h_\infty({\bf x};k) = 0$
in $x > R$ except for the exponential tail.
$R$ is called
the interaction range.
$H(p;k)$ can be obtained by $h_\infty({\bf x};k)$
using
the Fourier transformation,
\begin{eqnarray}
 H(p;k)
 &=& - \int_{-\infty}^\infty \!\!\! d^3 x \, h_\infty({\bf x};k) e^{-i {\bf p} \cdot {\bf x}}.
\label{eqn:H_pk_infinite_volume}
\end{eqnarray}
It is noted Eq.~(\ref{eqn:H_pk_infinite_volume}) at $p=k$
was employed for the relation between $\delta(k)$ and $\phi_\infty({\bf x};k)$ in $x > R$
in Ref.~\cite{Aoki:2005uf}.

The same S-wave amplitude $H(p;k)$ can be obtained from 
a reduced BS wave function on a finite volume $h({\bf x};k)$ 
with periodic boundary conditions as
\begin{eqnarray}
 H(p;k)
 &=& - F(k,L) \int_{-L/2}^{L/2} \!\! d^3 x \, h({\bf x};k) j_0(p x),
 \label{eqn:H_pk_finite_volume}
\end{eqnarray}
where
$L$ is the spatial extent.
$h({\bf x};k) = h_\infty({\bf x};k)/F(k,L)$ 
is evaluated from the BS wave function on the finite volume
$\phi({\bf x};k)$ as
$h({\bf x};k) = (\sum_{i=1}^{3} \partial_i^2 + k^2) \phi({\bf x};k)$.
The exponential factor $e^{-i{\bf p}\cdot{\bf x}}$ in 
Eq.~(\ref{eqn:H_pk_infinite_volume})
becomes
its $l=0$ component
of the spherical Bessel function
$j_0(p x)$ in Eq.~(\ref{eqn:H_pk_finite_volume}),
as
we consider only the S-wave scattering.
$F(k,L)$ is the finite volume correction of the two-pion state,
known as the Lellouch and L\"uscher factor~\cite{Lellouch:2000pv}.
A sufficient
condition of Eq.~(\ref{eqn:H_pk_finite_volume}) is $R < L/2$
on the finite volume.
If the condition is satisfied
and the exponential tail is negligible in the statistical precision,
we can change the range of the integration
from Eq.~(\ref{eqn:H_pk_infinite_volume}) to 
Eq.~(\ref{eqn:H_pk_finite_volume}), since $h({\bf x};k)$ in $x > R$
does not contribute to both
integrations.
This condition is
also required in
the finite volume method~\cite{Luscher:1986pf,Luscher:1990ux}.

Using Eqs.~(\ref{eqn:H_kk_infinite_volume}) and (\ref{eqn:H_pk_finite_volume})
at $p = k$ and removing overall factors including $F(k,L)$ by 
taking a ratio,
we can extract $\delta(k)$
from
$h({\bf x};k)$,
i.e.,
$\phi({\bf x};k)$ inside the interaction range.
It is in contrast to the finite volume method,
which was derived from $\phi({\bf x};k)$ outside 
the interaction range~\cite{Luscher:1990ux,Lin:2001ek,Aoki:2005uf}.
We present
Eq.~(\ref{eqn:H_pk_finite_volume}) 
can
be another method 
to calculate $\delta(k)$
and
the half-off-shell amplitude.


\vspace*{2mm}
{\bf \em Calculation of scattering amplitude on lattice:}
The $I=2$ two-pion BS wave function on the lattice $\phi({\bf x};k)$ 
is defined by
\begin{equation}
\phi({\bf x};k) = \langle 0 | \Phi({\bf x},t) | \pi^+ \pi^+, E_k \rangle 
e^{E_kt},
\end{equation}
where $| \pi^+ \pi^+, E_k \rangle$ is a ground state of two pions
in
the finite volume
and
$\Phi({\bf x},t) = \sum_{{\bf r}} 
\pi^+(R_{A_1^+}[{\bf x}] + {\bf r}, t) \pi^+({\bf r}, t)$
with a pion interpolating operator 
$\pi^{+}({\bf x},t) = \bar{d}({\bf x},t) \gamma_5 u({\bf x},t)$.
We perform $A_1^+$ projection $R_{A_1^+}[{\bf x}]$ to
attain
an S-wave scattering state.
We assume higher angular momentum scattering states of $l \geq 4$
are negligible 
compared to the ground state.

$\phi({\bf x};k)$ is derived
from a pion four-point function
in the center-of-mass frame
$C_{\pi\pi}({\bf x},t)$,
which
is given by
\begin{equation}
C_{\pi\pi}({\bf x},|t_{\rm sink}-t_{\rm src}|) =
 \langle 0 |
   \Phi({\bf x},t_{\rm sink}) \Omega(t_{\rm src}) 
 | 0 \rangle ,
\end{equation}
where 
$\Omega(t) = \sum_{{\bf x}_1,{\bf x}_2}\pi^+({\bf x}_1,t)\pi^+({\bf x}_2,t)$.
$t_{\rm sink}$ and $t_{\rm src}$ are the sink and source time slices,
respectively.
In a large $t = |t_{\rm sink} - t_{\rm src}| \gg 1$
region,
where the ground two-pion
contribution
dominates
$C_{\pi\pi}({\bf x},t)$,
$\phi({\bf x};k)$
can be obtained as
\begin{equation}
 C_k \phi({\bf x};k) = C_{\pi\pi}({\bf x},t) e^{E_k t},
\label{eqn:def_phixk_lat}
\end{equation}
with a constant $C_k$.

The reduced BS wave function on the lattice $h({\bf x};k)$
is
determined
by
$\phi({\bf x};k)$ as
\begin{equation}
 C_k h({\bf x};k) = (\Delta + k^2) [C_k \phi({\bf x};k)] ,
\label{eqn:def_hxk_lat}
\end{equation}
where $\Delta f({\bf x}) = \sum_{i=1}^{3} (f({\bf x}+\hat{i})+f({\bf x} - \hat{i})-2f({\bf x}))$.
The counterpart of Eq.~(\ref{eqn:H_pk_finite_volume}) on the lattice 
is given by
\begin{equation}
H_L(p;k) = -\sum_{{\bf x} \in L^3} C_k h({\bf x};k) 
j_0(px).
\label{eqn:H_L_pk_lattice}
\end{equation}
At $p=k$, we obtain the on-shell amplitude $H(k;k)$ as
\begin{equation}
 H(k;k) = H_L(k;k) / C_{00},
\label{eqn:H_L_kk_lattice}
\end{equation}
where $C_{00} = C_k / F(k,L)$.

The conventional finite volume method utilizes
two asymptotic forms of $\phi({\bf x};k)$ in $x > R$~\cite{Luscher:1990ux,Aoki:2005uf},
\begin{eqnarray}
C_k \phi({\bf x};k) &=& v_{00} G({\bf x};k) 
\label{eqn:expand_phi_G}\\
&=& C_{00} e^{i\delta(k)} \sin(kx+\delta(k)) / kx + \cdots ,
\label{eqn:expand_phi_sin}
\end{eqnarray}
where 
$G({\bf x};k) = \sum_{{\bf p} \in \Gamma} e^{i{\bf x}\cdot{\bf p}}(p^2 - k^2)^{-1}/L^3$
with $\Gamma = \{{\bf p} | {\bf p} = (2\pi/L){\bf n}, {\bf n} \in {\bf Z}^3\}$ 
being the Green function on the finite volume $L^3$ 
with the periodic boundary condition.
$v_{00}$ and $C_{00}$ are constants.
The dots express terms of
the spherical Bessel function
$j_l(kx)$ of $l \ge 4$.
We can expand $G({\bf x};k)$ by $j_l(kx)$ and
$l=0$ Neumann function $n_0(kx)$. Comparing
the coefficients of $j_0(kx)$ and $n_0(kx)$
in the expanded form of Eq.~(12) with those of Eq.~(13)
yields two equations,
\begin{eqnarray}
 C_{00} H(k;k) & = & v_{00} ,
\label{eqn:finite_volume_1}\\
k \cot\delta(k) C_{00} H(k;k) &=& 4 \pi v_{00} g_{00}(k) ,
\label{eqn:finite_volume_2}
\end{eqnarray}
where $g_{00}(k) = \sum_{{\bf p} \in \Gamma} (p^2 - k^2)^{-1}/L^3$.
Taking a ratio of these equations
gives the formula of the finite volume method
to evaluate the S-wave $\delta(k)$,
$k\cot\delta(k) = 4\pi g_{00}(k)$.
Using Eq.~(\ref{eqn:finite_volume_1}), we will examine
the consistency
between a numerical result from $H(k;k)$ and that from
the finite volume method.


\vspace*{2mm}
{\bf \em Setup:}
Our simulation is executed in the quenched lattice QCD.
The simulation setup conforms to Refs.~\cite{Aoki:2005uf,AliKhan:2001tx}.
The gauge action is Iwasaki type~\cite{Iwasaki:2011jk, *Iwasaki:1985we}.
Gauge configurations are generated at the bare coupling $\beta = 2.334$
on the lattice size of
$24^3 \times 64$
by the Hybrid Monte Carlo algorithm.
The number of configurations is 200, separated by 100 trajectories.
The lattice spacing has been found to be $a^{-1} = 1.207$ GeV~\cite{AliKhan:2001tx}.
The quark action is a mean field improved Clover type~\cite{Sheikholeslami:1985ij}
with $C_{\rm SW} = 1.398$.
The quark hopping parameter is $\kappa_{\rm val} = 0.1340$,
corresponding to
$m_\pi = 0.85755(25)$ GeV.

We employ a random $Z(2)$ source at $t_{\rm src}$
spread in the spatial volume 
and also in the spin and color spaces to reduce the calculation cost.
We use four random sources at one time slice, and
calculate six $C_{\pi\pi}({\bf x},t)$ from all the possible 
combinations
of
the two quark propagators with the different
random sources.
We repeat the calculation every two time slices
on
a configuration
to increase the statistics.
We also employ a wall source for a check of source independence of our results.
Wall sources are set at $t_{\rm src}$ and $t_{\rm src} + 1$ to avoid Fierz
contamination. The total number of the wall source points is 16.
The quark propagators are solved with the periodic boundary condition
in space and with the Dirichlet boundary condition 
in time, which is imposed
to be
separated by 12 time slices from $t_{\rm src}$.


\begin{figure}[!tb]
 \centering
 \includegraphics[width=0.9\columnwidth]{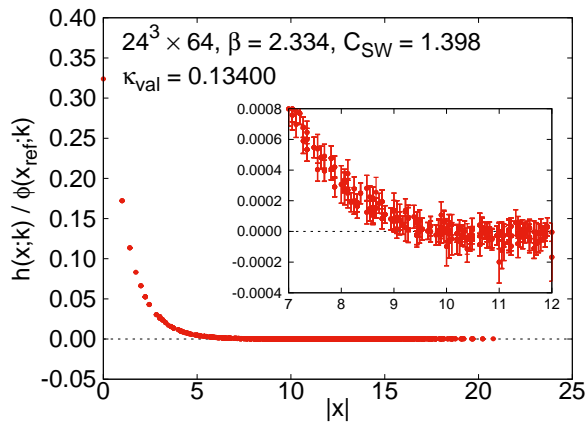}
 \caption{
  \label{fig:reduced_phi_L}
  $h({\bf x};k)/\phi({\bf x}_{\rm ref};k)$
  with ${\bf x}_{\rm ref} = (12,7,2)$
  as
  a function of $x$.
  The inside panel shows the same data with
  an
  enlarged scale
  in $7 \le x \le 12$.
 }
\end{figure}

\vspace*{2mm}
{\bf \em Result:}
Figure~\ref{fig:reduced_phi_L} illustrates
the result of $h({\bf x};k)/\phi({\bf x}_{\rm ref};k)$
as
a function of $x$,
which is calculated from a ratio of Eq.~(\ref{eqn:def_hxk_lat})
to Eq.~(\ref{eqn:def_phixk_lat}) at ${\bf x}_{\rm ref} = (12,7,2)$.
In the figure and the following analyses,
we choose $t=44$
for
$C_k \phi({\bf x};k)$ in Eq.~(\ref{eqn:def_phixk_lat}).
We use $k^2$ determined from $E_k$
by
a single exponential fit of
$C_{\pi\pi}(t) = \sum_{\bf x} C_{\pi\pi}({\bf x},t)$
with a fit range of
$12 \le t \le 44$.
The result
of
$k^2$ is presented in
Table~\ref{tab:k2}.
In $x > 10$, $h({\bf x};k)/\phi({\bf x}_{\rm ref};k)$ becomes 
consistent with zero in our statistics,
suggesting
the interaction range
$R \sim 10 < L/2 = 12$.
Our calculation satisfies the
sufficient
condition to use 
Eq.~(\ref{eqn:H_pk_finite_volume}).
Our value of $R$ is consistent with
that
in Ref.~\cite{Aoki:2005uf}.

\begin{table}[tb]
\begin{center}
\begin{tabular}{ccc}\hline\hline
$k^2$[GeV$^2$] & $a_0/m_\pi(E_k)$[GeV$^{-2}$] & $a_0/m_\pi(H(k;k))$[GeV$^{-2}$] \\ \hline
1.549(45)$\times 10^{-3}$ & $-0.997(27)$ & $-0.994(25)$ \\ \hline \hline 
\end{tabular}
 \caption{
 Results for $k^2$, $a_0/m_\pi(E_k)$ obtained by the finite volume method
 and $a_0/m_\pi(H(k;k))$ obtained by Eq.~(\ref{eqn:delta_HLkk}).}
  \label{tab:k2}
\end{center}
\end{table}

$H_L(k;k)$ is evaluated by performing the spatial summation
in Eq.~(\ref{eqn:H_L_pk_lattice}) with $p = k$.
The result by all spatial summation
agrees with that by up to $|{\bf x}| = 10 \sim R$,
indicating the estimate of $R$ is valid.

\begin{figure}[!t]
 \centering
 \includegraphics[width=0.9\columnwidth]{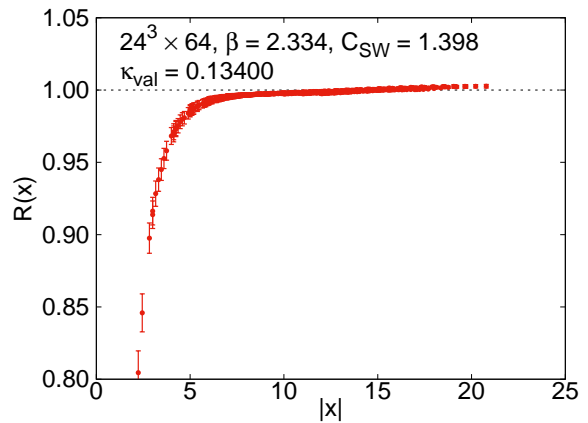}
 \caption{
  \label{fig:test_hkk}
  $R({\bf x})$ defined in Eq.~(\ref{eqn:test_Hkk_lattice})
  as
  a function of $x$. The dashed line expresses $R({\bf x})=1$.
 }
\end{figure}

We also examine
whether $H_L(k;k)$ calculated on the lattice satisfies
the condition of the finite volume method of Eq.~(\ref{eqn:finite_volume_1}).
An indicator
quantity $R({\bf x})$
is defined
as
\begin{equation}
 R({\bf x}) = 
 \frac{H_L(k;k)}
      {C_k \phi({\bf x};k)}
 G({\bf x};k)
 \xrightarrow[x > R]{}
 \frac{C_{00} H(k;k)}{v_{00} G({\bf x};k)}
 G({\bf x};k) ,
\label{eqn:test_Hkk_lattice}
\end{equation}
where we use
Eqs.~(\ref{eqn:H_L_kk_lattice}) and
(\ref{eqn:expand_phi_G}) in the arrow.
$G({\bf x};k)$ is evaluated using the formula in 
Appendix~B of Ref.~\cite{Aoki:2005uf}.
If Eq.~(\ref{eqn:finite_volume_1}) is satisfied, 
$R({\bf x})$
equals
unity.
The result of $R({\bf x})$
is plotted in Fig.~\ref{fig:test_hkk}.
$R({\bf x})$
approaches
unity in a large $x$ region,
as expected.
In $x > R \sim 10$,
$R({\bf x})$ agrees with unity
within 2 standard deviations.

$\tan \delta(k)$ is evaluated from $H_L(k;k)$
using
the
asymptotic form of $C_k \phi({\bf x},t)$ in 
Eq.~(\ref{eqn:expand_phi_sin}) at
a
reference point ${\bf x}_{\rm ref} = (12,7,2)$.
We
choose ${\bf x}_{\rm ref}$, examining
the size of the
leading $l=4$ contribution in the dots terms
by $Y_{40}(R_{A_1^+}[{\bf x} / x]) j_4(k x) / (Y_{00}(R_{A_1^+}[{\bf x} / x]) j_0(k x))$
at each position in $x > R$,
where $Y_{lm}({\bf x} / x)$ are spherical harmonics.
${\bf x}_{\rm ref}$ is chosen such that
$|Y_{40}(R_{A_1^+}[{\bf x} / x]) j_4(k x) / (Y_{00}(R_{A_1^+}[{\bf x} / x]) j_0(k x))| < 10^{-6}$.
$\tan \delta(k)$ is
then
given by
\begin{equation}
 \tan \delta(k)
 = \frac{\sin(k x_{\rm ref})}{4 \pi x_{\rm ref} C_k \phi({\bf x}_{\rm ref};k)/H_L(k;k)
 - \cos(k x_{\rm ref})} ,
\label{eqn:delta_HLkk}
\end{equation}
where $H_L(k;k)/(C_k \phi({\bf x}_{\rm ref};k)) = 4 \pi x_{\rm ref} \sin \delta(k) / \sin(k x_{\rm ref} + \delta(k))$.
In the ratio $H_L(k;k)/(C_k \phi({\bf x}_{\rm ref};k))$,
the overall factors of $H_L(k;k)$, $C_{00} e^{i \delta(k)}$,
are canceled.
We cannot determine the phase $e^{i \delta(k)}$ by this method.
Correspondingly, the determination is impossible
by the finite volume method.

\begin{figure}[!tb]
 \centering
 \includegraphics[width=0.9\columnwidth]{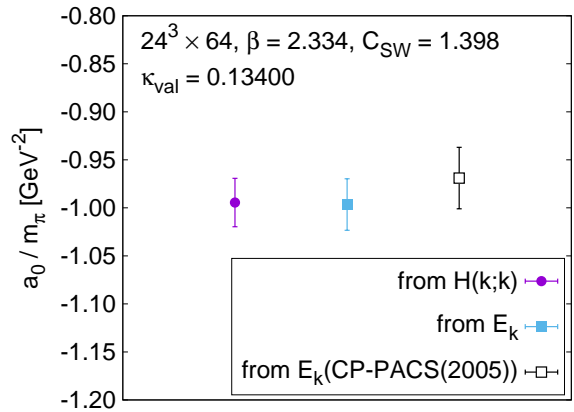}
 \caption{
  \label{fig:comparison}
  Comparison of scattering lengths $a_0 / m_\pi$ from
  $H(k;k)$ with ${\bf x}_{\rm ref} = (12,7,2)$ (circle)
  and from the finite volume method (square),
  together with the result of Ref.~\cite{Aoki:2005uf}
  (open square).
 }
\end{figure}

The scattering length $a_0$
is
obtained from $\tan \delta(k)$ as
$a_0 / m_\pi = \tan \delta(k) / (k m_\pi)$.
We omit higher-order terms of $k^2$.
$k^2$ is small enough in our calculation.
Our result from Eq.~(\ref{eqn:delta_HLkk}) agrees with that
from the finite volume method
as shown in Table~\ref{tab:k2}
and the value
in Ref.~\cite{Aoki:2005uf}.
Those results are compared in Fig.~\ref{fig:comparison}.
It is noted ambiguity from the choice of ${\bf x}_{\rm ref}$ is
well below the statistical error.
For example, we have $a_0 / m_\pi = -1.001(27)$ with ${\bf x}_{\rm ref} = (10,4,4)$
and $a_0 / m_\pi = -1.006(27)$ with ${\bf x}_{\rm ref} = (9,5,2)$.

As a check,
the largest $l = 4$ contribution in the dots terms
of Eq.~(\ref{eqn:expand_phi_sin}),
which appears
at ${\bf x} = (L/2,L/2,L/2)$,
is also estimated.
Using this position as ${\bf x}_{\rm ref}$ still leads to
a similar result, $a_0 / m_\pi = -0.967(25)$.
It confirms
the systematic error
from the $l = 4$ contribution
is not significant compared to the statistical error.

The analysis using Eq.~(\ref{eqn:test_Hkk_lattice})
and
the comparison of $a_0 / m_\pi$
conclude
$H_L(k;k)$ calculated on the lattice
satisfies
Eq.~(\ref{eqn:finite_volume_1})
and gives $\delta(k)$ consistent
with
that by
the finite volume method.
The uncertainties in the two methods are also comparable.
It should be emphasized that
$H_L(k;k)$ is calculated using $\phi({\bf x};k)$ inside
the interaction range,
in contrast to 
the derivation of the finite volume method
and the analysis of Ref.~\cite{Aoki:2005uf} using $\phi({\bf x};k)$ in $x > R$.

\begin{figure}[!tb]
 \centering
 \includegraphics[width=0.9\columnwidth]{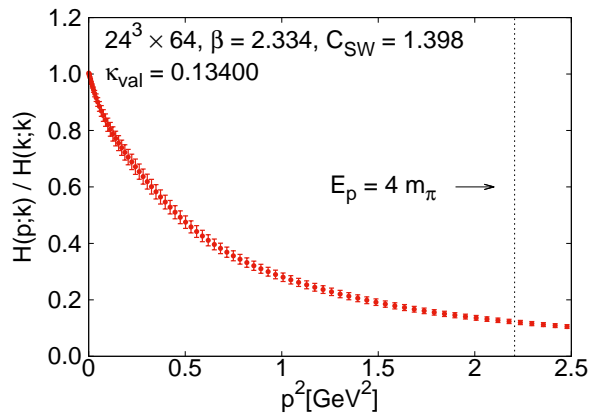}
 \caption{
  \label{fig:half_off_shell}
  Half-off-shell amplitude $H(p;k)$ normalized by the on-shell value $H(k;k)$
  as
  a function of $p^2$. The vertical dotted line denotes the threshold
  energy of the two-pion scattering.
 }
\end{figure}

We also
evaluate the half-off-shell amplitude $H(p;k)$ using 
Eq.~(\ref{eqn:H_L_pk_lattice}).
Figure~\ref{fig:half_off_shell} presents $H(p;k)$
as
a function of $p^2$
normalized by its on-shell value as
$\frac{H_L(p;k)}{H_L(k;k)} = \frac{H(p;k)}{H(k;k)}$.
A clear signal of $H(p;k)$ is obtained.
The result decreases as $p^2$ increases.
In the figure,
the
inelastic
threshold of the two-pion scattering is also plotted.
It is smooth at the threshold,
which may be
due to the quenched approximation.
While $H(p;k)$ cannot be directly compared with experiment,
it might be
an additional input to
constrain parameters of effective models of
hadron scatterings 
as a supplement to experimental data.

It is noted that the half-off-shell amplitude on the lattice itself depends on the choice of the
operator. The dependence, however, is canceled in a ratio of $H_L(p;k) / H_L(k;k)$.
We numerically confirmed the operator independence of $H_L(p;k) / H_L(k;k) = H(p;k) / H(k;k)$
by the use of wall sources, in addition to random sources.
Both results agree within errors.
If we employ a sink smearing of one of the pion operators for the BS wave function,
an additional overall factor appears.
This additional factor can be analytically erased~\cite{Kawai:2017goq}.

We further
attempt to extract information of the scattering from $H(p;k)$
with two assumptions.
We assume that around $p^2 = k^2$
the phase of $H(p;k)$ equals $e^{i \delta(k)}$, and
$\partial (H(p;k) e^{-i \delta(k)}) / \partial p^2 \sim \partial (H(p;p) e^{-i \delta(p)}) / \partial p^2$.
Under the assumptions,
the effective range expansion
$k \cot \delta(k) = a_0^{-1} + r k^2 + O(k^4)$
leads to an estimate of the effective range $r$,
\begin{equation}
 r = - \frac{2 k^2
             H^\prime
             + \sin^2 \delta(k)}
            {2 k \sin \delta(k) \cos \delta(k)} ,
\end{equation}
where
$H^\prime$ is 
the slope of
$H(p;k) / H(k;k)$
with respect to $p^2$ at $p^2 = k^2$.
Using our measured data,
we estimate $r = -2.64(41)$ GeV$^{-1}$,
which is not inconsistent with
$r = -0.3(8.4)$ GeV$^{-1}$ using the data
of $k \cot \delta(k)$ in Ref.~\cite{Aoki:2005uf}.

\vspace*{2mm}
{\bf \em Summary:}
We proposed a new method to calculate
the scattering phase shift $\delta(k)$
from
the on-shell scattering
amplitude
on the lattice $H_L(k;k)$
obtained with the
BS wave function $\phi({\bf x};k)$ inside 
the interaction range
by definition of
quantum field theory.
By a simulation of S-wave $I=2$ two-pion scattering
in the center-of-mass frame,
the consistency of our method with the finite volume method
was examined.
Our data
of $R({\bf x})$ defined by Eq.~(\ref{eqn:test_Hkk_lattice})
were found to satisfy the condition in the finite volume method of
Eq.~(\ref{eqn:finite_volume_1})
in $x > R$.
Our result of the scattering length $a_0 / m_\pi$ from $H_L(k;k)$
agrees
with the
value
from the finite volume method.
The consistency
concludes
our method
using information inside the interaction range
can be an alternative
to
the finite volume method
using data outside
the
interaction range.

We note issues of scaling violation. One is rotational symmetry breaking of $h({\bf x};k)$.
It can be considered as a scaling violation of $H_L(k;k)$.
The size of the rotational symmetry breaking in $H_L(k;k)$ is estimated by the difference
between $H_L(k;k)$ with the minimum and maximum values of $h({\bf x};k)$ at each
degenerate point of $|{\bf x}|$. The breaking effect is found 
to be 3\%, close to our statistical error.
Another is lattice artifacts of $h({\bf x};k)$ at small $|{\bf x}|$.
The artifacts are expected to be significant,
but suppressed in $H_L(k;k)$. It is clearly understood
by the Jacobian factor $r^2$ of $H(p;k)$ in Eq.~(\ref{eqn:H_pk_finite_volume})
in spherical coordinates.
We noticed agreement between $a_0/m_\pi$ from $H(k;k)$
and that from the finite volume method
in Fig.~\ref{fig:comparison}
indicates each method has a similar size of the scaling violation.
Nevertheless, it is important future work to perform simulations at
a different value of the lattice spacing for the investigation of
the scaling violation.

We
also
evaluated the half-off-shell 
scattering amplitude $H_L(p;k)$ by lattice QCD.
It might be a supplemental input to theoretical models of hadrons.
We
extracted
the effective range
from $H_L(p;k)$
with some assumptions.
Although it was not inconsistent
with the result
from Ref.~\cite{Aoki:2005uf},
our assumptions still need to be validated.

We remark
it is essential to obtain the reduced BS wave function $h({\bf x};k)$ 
for
the on-shell and half-off-shell amplitudes.
$h({\bf x};k)$ is directly related to the amplitudes $H_L(p;k)$
in
a simple form of Eq.~(\ref{eqn:H_L_pk_lattice}).
We may derive similar relations between
the scattering amplitude and the reduced BS wave function in 
moving frames and scattering systems of more than two particles.


\vspace*{2mm}
{\bf \em Acknowledgments:}
We thank N.~Ishizuka and Y.~Kuramashi for their careful reading of the manuscript and useful comments.
Our simulation was performed on COMA
under Interdisciplinary Computational Science Program of
Center for Computational Sciences, University of Tsukuba.
This work is in part based on the Bridge++ code~\cite{Bridge}.
This work is supported in part by JSPS KAKENHI Grants
No. 15K05068 and No. 16H06002.

\bibliographystyle{apsrev4-1}
\bibliography{reference}

\end{document}